\documentstyle[12pt,graphicx]{article}
\title{ The Largest Scale We Can Detect in the Universe and the
Inflation }
\author{\small Wen Zhao, Yang Zhang \\
        \small Astrophysics Center \\
        \small University of Science and Technology of China \\
        \small Hefei, Anhui, China }
 \date{}

\begin{document}
\maketitle
\baselineskip=19truept
\def\vek{\vec{k}}

\newcommand{\be}{\begin{equation}}
\newcommand{\ee}{\end{equation}}
\newcommand{\ba}{\begin{eqnarray}}
\newcommand{\ea}{\end{eqnarray}}

\sf
\small

\begin{center}
\Large  Abstract
\end{center}
\begin{quote}
 {\small
 From the damping of the Cosmic Microwave Background Radiation
 (CMB) anisotropy power spectrum at large scale and the recent
 accelerating expansion of the Universe,
 we find that, there may be a largest scale which we can detect
 in the Universe. From this, we can get the inflation parameters
 as spectrum index $n_s$ of the initial scalar spectrum, e-fold
 $N$, Hubble parameter $H$, the ratio of tensor and scalar $r$,
 the lasting time of reheating stage $\alpha$ for special inflation
 models.  We do them in three inflation models, and find that all
 the results fit fairly well with the observations and the inflation
 theory.

 }
\end{quote}



e-mail: wzhao7@mail.ustc.edu.cn

\newpage
\baselineskip=19truept

\begin{center}
\end{center}

The release of the high resolution full sky Wilkinson Microwave
Anisotropy Probe (WMAP) data shows that the data is consistent
with the predictions of the standard inflation-$\Lambda$CDM cosmic
model, expect for several
puzzles\cite{net}\cite{map1}\cite{Map1}\cite{Map2}\cite{map2}\cite{map4}\cite{map3}.
One of those is that the power spectrum shows that there is a much
damping at the large scale
$l<10$\cite{map1}\cite{map4}\cite{map3}, which has been much
deeply discussed\cite{damp}. The simplest explanation is that
there is a cut at the very small wavenumber (IR cut-off) for the
initial scalar perturbation. We show the power spectrum of CMB in
figure (1), where the dots are the WMAP observation result, and
the red line is the spectrum without IR cut-off, the green line
with cut-off at 1/k=4000Mpc, the blue line with cut-off at
1/k=3000Mpc, the magenta line with cut-off at 1/k=2000Mpc. We have
set the scale factor of now $a_0=1$, and chosen the cosmological
parameters as $\Omega_m=0.047$, $\Omega_{\Lambda}=0.693$,
$\Omega_{dm}=0.29$, $h=0.72$, $n_s=0.99$\cite{map2}, and without
consider the reionization and the running of $n_s$. From this
figure, we find that the cut of the initial perturbation is nearly
at $1/k\simeq2000-4000Mpc$, but why this cut-off exist?

We all know that there is an inflation stage at the very early
Universe\cite{kolb}\cite{lid}in the inflation-$\Lambda CDM $
cosmic model. The attraction of this paradigm is that it can set
the initial conditions for the subsequent hot big bang, which
otherwise have to be imposed by hand. One of these is that there
be no unwanted relics (particles or topological defects which
survive to the present and contradict observation). Another is
that the initial density parameter should have the value
$\Omega=1$ to very accuracy, to ensure that its present value has
at least roughly this value. There is also the requirement that
the Universe be homogeneous and isotropic to high accuracy. But
the most important is that the scale-invariant initial scalar
perturbation power spectrum which predicted, has been detected
from CMB and LSS, especially the recent
WMAP\cite{map1}\cite{map2}\cite{map3} and SDSS\cite{sdss1}
observation.

In this paradigm, the scale factor expanded much more rapidly in
the initial inflation stage than the horizon\footnote{The horizon
in this paper all denotes the particle horizon.}. From the sketch
figure (2), we find that the scale of $a/k_1$ goes out the horizon
at inflation, and reenteres the the horizon at radiation
(dust)-dominating stage. Where $k_0$ is the smallest wavenumber,
which had been in the horizon, when $k<k_0$, the wave had never
been in the horizon, so the initial power spectrum should only at
$k>k_0$. So for the scale with $l_2=2\pi a_i/k_2$ is larger that
$l_H=2\pi a_i/k_0$ ( $a_i$ is the scale factor at $t_i$, $k_0$
satisfies that $a_i/k_0=1/H$ , where $H$ is the Hubble parameter
of inflation), the Universe will never be homogeneous and
isotropic. if this cut of $k_0$ is just nearly $1/k_0\sim3000Mpc$,
it will naturally answer the damping of the CMB power spectrum.
The discussion in below will all base on this idea. If the
Universe was dominated by dust, the scale factor $\propto
t^{2/3}$, but the horizon is nearly $\propto ct$ (c is the speed
of light), so there must be a time $t_m$, $2\pi a_m/k_0\sim t_m$,
after which, the Universe will not be homogeneous and isotropic.
But that was not the true Universe.

A wide range of observational evidence indicates that our Universe
may be accelerating expansion\cite{sne}
\cite{sdss1}\cite{sdss2}\cite{map1}\cite{map2}\cite{cmb}\cite{age}.
If we assume that long-range gravity obeys Einstein's General
Relativity, this suggests that most of our Universe is in some
form of smooth dark energy with $\omega\simeq-1$, which can
comprise $\sim 70\%$ of the critical energy. In this accelerating
expansion Universe, there must be a time $t_a$, after which the
scale factor would expand fast than the horizon. At this point,
the wave which satisfied that $2\pi a_a/k_a=l_a$ (where $a_a$ is
the scale factor , and $l_a$ is the horizon at $t_a$), exactly
reenter the horizon, and immediately went out, so this is the
largest scale which can reenter the horizon. It is also the
largest scale which we can detect in the Universe.

For the flat Robertson-Walker metric \be
ds^2=-c^2dt^2-a^2(dx^2+dy^2+dz^2),\ee the time $t_a$ is nearly the
time when the Universe began to accelerating expansion, so
$d^2a/dt^2=0$ at time $t=t_a$, which means that $\omega=\Sigma
p_i/\Sigma\rho_i=-1/3$. For now the Universe is made of
$\Omega_m=0.047$, $\Omega_{dm}=0.29$, $\Omega_{de}=0.693$, and
$\omega_{de}=-1$, one can get $z_1=0.653$. The particle horizon at
time $t$ is defined as: \be l(t)=a(t)\int_0^t
\frac{dt'}{a(t')},\ee where the scale factor $a$ can be get from
the Friedmann equations. From which one can get the horizon at
this time is $l_a=6173Mpc$, where we have used the the Hubble
constant $h=0.72$. So $1/k_a=1625Mpc$. We are surprised to find
that $k_a$ is nearly equate to $k_0$, which we have get from
before. We know $1/k_0$ is the largest scale which had been in the
horizon at inflation stage, so only if when the scale is smaller
than which, the Universe can keep homogeneous and isotropic. But
$1/k_1$ is the largest scale which can reenter the horizon in the
accelerating Universe, so if the Universe is always homogeneous
and isotropic, which requires that $k_0\leq k_a$ (from before we
know that it is satisfied), so we know that this accelerating
expansion Universe will always be homogeneous and isotropic even
if the inflation only expanded a period of time. This is one of
the most important difference between the accelerating Universe
and the decelerating Universe. If we can accept the below
Assumption: assume the exact state that $k_0\equiv k_a=k_0$ (from
before, we find $k_0$ is a litter smaller than $k_a$, which is for
we considering the inflation and dark energy in a crud way with
$H$=constant, $\Omega_{de}=-1$, which all can affect $k_0$ and
$k_a$ to form this difference between them), which will mean that
$1/k_a$ is always the largest scale we can detect in the Universe.
The evolution of scale factor and horizon with time are shown in
the sketch figure 2, where the black (solid) line is the particle
horizon, the red (dash) line denotes the evolution of $a/k_1$, the
blue (dot) line denotes $a/k_0$, and the magenta (dash dot) line
denotes $a/k_3$, where $k_1>k_0>k_2$. We will discuss if this
Assumption is rational in below.



With this Assumption, we can get the inflation parameters: the
index of initial spectrum $n_s$, e-fold $N$, Hubble parameter of
inflation $H$, the ratio of tensor and scalar perturbation $r$,
and the expansion times of the scale factor in the reheating stage
$\alpha$ for the special inflation models.

For \be r=\frac{\Delta_h^2}{\Delta_R^2}, \ee where the scalar
perturbation power spectrum is\cite{map3}\cite{map2} \be
\Delta_R^2=2.95\times10^{-9}A, \ee  The tensor power
spectrum\cite{tensor}
\be\Delta_h^2=C^2(\mu)\frac{4}{\pi}\frac{H^2}{m_p^2},\ee one can
get the first relation \be
\frac{H}{m_p}=(\frac{2.95\times10^{-9}Ar}{4C^2/\pi})^{1/2}.
\label{1}\ee where $A$ is the scalar power spectrum amplitude,
which is nearly $0.9$ from WMAP result, and the constant number
$C(\mu)\simeq1$, $m_p=2.4\times10^{18}GeV$ is the Planck energy
scale.

If the inflation ended at time $t_e$. The expansion times of the
scale factor between $t_i$ and $t_e$ is $a_e/a_i=e^N$, where $N$
is the e-fold of the inflation. We can easily get another formula
$a_a/a_e=\alpha H/T_a$, where $T_a$ is the CMB temperature at time
$t_a$, $\alpha$ was defined as: \be \alpha=a_{r}/a_{re}, \ee where
$a_{r}$ and $a_{re}$ are the scale factor at the end of the
reheating stage and the beginning of the reheating stage. Here we
have used that $aT=constant$ in the Universe. From the Assumption
as before, one can get the second relation: \be e^N=\alpha
H/T_1=2.2\times10^{12}\alpha (H/GeV). \label{2}\ee

From $a_i/k_i=1/H$, $2\pi a_1/k_1=l_1$, and $k_i=k_1$, one can get
the third relation: \be
a_1/a_i=e^{2N}=l_1/l_H=2.1\times10^{42}(H/GeV).\label{3}\ee

From the relation of (\ref{2}) and (\ref{3}), one can get a simple
constraint on the Hubble parameter of the inflation: \be
H<4.3\times10^{17}GeV\equiv H_m,\ee for $\alpha>1$, which is only
dependent on the Assumption as before but not on the inflation
models. For the special inflation cosmic models, one can get
another two relations about $N$ \& $n_s$ and $r$ \& $n_s$, here we
consider three inflation models:

(1) $\lambda\phi^4$ inflation model:
 In this model, we have two simple relations of
 \be
 N=\frac{3}{1-n_s},~~~~~~r=\frac{16}{3}(1-n_s),
 \ee
 combining which with the relations (\ref{1}), (\ref{2}), (\ref{3}),
 one can get the inflation parameters as below:
 \be
 n_s=0.95,~~N=64.53,~~r=0.25,~~\alpha=88.42,~~H=5.46\times10^{13}GeV<H_m;
 \ee

 (2) $V(\phi)=\Lambda^4(\phi/\mu)^2$ inflation model:
 In this model, two simple relations are
 \be
 N=\frac{2}{1-n_s},~~~~~~r=4(1-n_s),
 \ee
 so we can get the inflation parameters in the same way:
 \be
 n_s=0.97,~~N=64.35,~~r=0.12,~~\alpha=105.55,~~H=3.86\times10^{13}GeV<H_m;
 \ee

(3) $V(\phi)=\Lambda^4(\phi/\mu)^4$ inflation model:
 In this model, the two simple relations are
 \be
 N=\frac{3}{1-n_s},~~~~~~r=\frac{16}{3}(1-n_s),
 \ee
 so the inflation parameters are:
  \be
 n_s=0.95,~~N=64.53,~~r=0.25,~~\alpha=88.42,~~H=5.46\times10^{13}GeV<H_m;
 \ee

From the observation, one gets the index of the initial spectrum
of $n_s=0.99\pm0.04$ from WMAP only, and $n_s=0.96\pm0.02$ from
WMAPext+2dFGRS+Lyman~$\alpha$\cite{map2}. The index from these
three models are all fit fairly well with the observation. The
constraint on $r$ form WMAP and SDSS is
$r<0.36$\cite{map3}\cite{sdss}, which is also consistent with the
calculation results. Naturally, we accept that $N\geq 60$,
$H\sim10^{13}GeV$\cite{map3}, which are all consistent with the
models results. We also get the $\alpha$ is nearly $100$
(independent on the reheating models), which tell us that the
reheating process is a very quick process, at this stage, the
scale factor only expanded nearly $100$ times. From the before
calculation, we find that, there inflation parameters are only
dependent on the two relations of $N$ \& $n_s$, and $r$ \& $n_s$.
The models of (1) and (3) have the same relations, so they give
the same inflation parameters.

In summary, from the damping of the CMB anisotropy power spectrum
at large scale, we think it is for that the initial power spectrum
has a cut-off at $k<k_0$, which generated for the inflation must
began at some time. One find that this cut-off is nearly equal to
the largest scale which can reenter the horizon in the
accelerating Universe. From this we elicit a simple Assumption:
$k_0\equiv k_a$ so the largest scale, which can be in the particle
horizon in the inflation-accelerating expansion Universe is $\sim
1/k_a$. This Assumption will keep that the Universe will always be
homogeneous and isotropic. To check the rationality of this
Assumption, we calculated the inflation parameters od $n_s$, $N$,
$H$,$\alpha$ in three inflation models from this Assumption, and
find that they are all consistent with the observation and the
inflation theory.

The accelerating expansion of the recent Universe is a very puzzle
for cosmologist. Here we consider this question in another point:
for the early inflation stage must have a beginning at some time,
which make that there exist a largest scale $1/k_0$, when the
scale is larger than it, the Universe will not be homogeneous and
isotropic. The Universe began to decelerating expansion after
inflation, if this deed was kept for all time, which will make the
Universe not be homogeneous and isotropic at large scale at some
time, for the expansion of the horizon is much quick that the
scale factor. But to our surprise is that the Universe began to
accelerating expansion at redshift $z\simeq0.653$, which exactly
elegantly make the Universe always be homogeneous and isotropic.
Usually, we always ask the question: why the accelerating
expansion exist? but here, we ask another question: why it is
needed? Our answer is that: to keep the cosmological principle
being right for all time. About why the Universe is this? which
however, need to research. But from the discussion above, we think
the below viewpoint is rational: the recent accelerating must have
some relation with the early inflation, these two accelerating
expansion stages make Universe always be homogeneous and
isotropic.

~

~

~

ACKNOWLEDGMENT: We thank B. Feng for helpful discussion, and also
thank T.Padmanabhan for his advise and his advised
paper\cite{pad}. We acknowledge the using of CMBFAST
program\cite{fast}. Y. Zhang's research work has been supported by
the Chinese NSF (10173008) and by NKBRSF (G19990754). W.Zhao's
work is partially supported by Graduate Student Research Funding
from USTC.

~

~

~

~

~


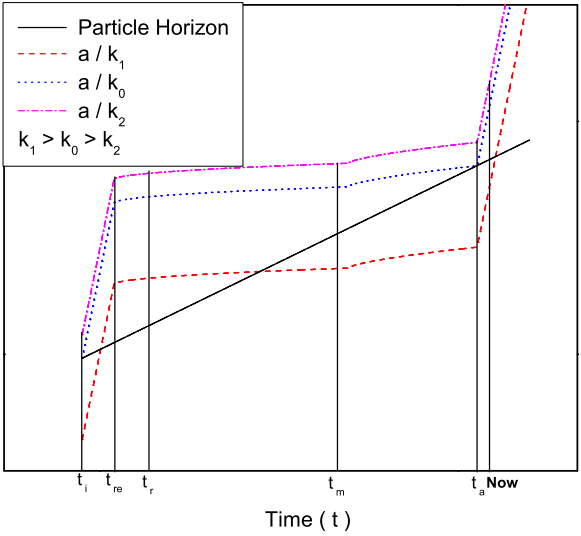
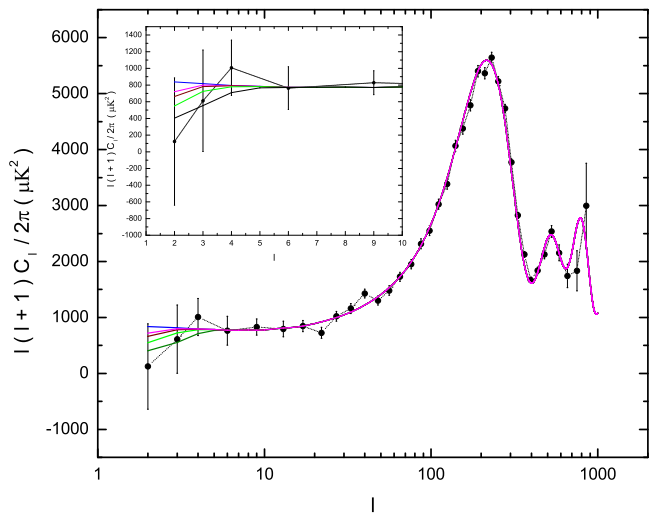
\end{document}